\RequirePackage{fix-cm}
\documentclass[smallextended]{svjour3} 
\smartqed 
\usepackage{graphicx}
\begin{document}

\title{ Massive Neutron Star Models with Parabolic Cores
}


\author{P S Negi 
}


\institute{Department of Physics, \at
Kumaun University, Nainital\\
\email{psneginainital63@gmail.com} 
}

\date{Received: date / Accepted: date}

\maketitle

\begin{abstract}
The results of the investigation of the core-envelope model presented in Negi et al. \cite{Ref1} have been discussed in view of the reference \cite{Ref2} . It is seen that there are significant changes in the results to be addressed. In addition, I have also calculated the gravitational binding energy, causality and pulsational stability of the structures which were not considered in Negi et al. \cite{Ref1} .
The modified results have important consequences to model neutron stars and pulsars. The maximum neutron star mass obtained in this study corresponds to the mean value of the classical results obtained by Rhodes \& Ruffini \cite {Ref3} and the upper bound on neutron star mass
obtained by Kalogera \& Byam \cite {Ref4} and is much closer to the most recent theoretical estimate made by Sotani \cite{Ref5}. On one hand, when there are only few equations of state (EOSs) available in the literature which can fulfil the recent observational constraint imposed by the largest neutron star masses around 2$M_\odot$\cite{Ref6}, \cite{Ref7}, \cite{Ref8}, the present analytic models, on the other hand, can comfortably satisfy this constraint. Furthermore, the maximum allowed value of compactness parameter $u(\equiv M/a$; mass to size ratio in geometrized units) $ \leq 0.30$ obtained in this study is also consistent with an absolute maximum value of $ u_{\rm max} = 0.333^{+0.001}_{-0.005}$ resulting from the observation of binary neutron stars merger GW170817  (see, e.g.\cite{Ref9}).
\keywords{Static Spherical Structures \and Analytic Solutions \and Neutron Stars}
\end{abstract}

\section{Introduction}
\label{intro}
The study carried out by Negi et al. \cite{Ref1} deals with the construction of a core-envelope model of static and spherical mass distribution characterized
by exact solutions of Einstein's field equations. The core of the model is described by Tolman's VII solution (TDR solution) matched smoothly at the core-boundary. The region of the envelope is
described by Tolman's V solution which is finally matched to vacuum Schwarzschild solution. The core-envelope boundary of the model is assured by matching of all the four variables - pressure ($P$), energy density ($E$) and both of the metric parameters $\nu$ and $\lambda$ with recourse to the computational method. The complete solutions with appropriate references for both the regions (the core and the envelope) are available in Negi et al. \cite{Ref1}. However, it appears that while computing the core-envelope boundary and other parameters by using equation (19) - (22) in Ref.[1] and thereafter following the expression for $w_b$, some error occurred in the computation of Negi et al. \cite{Ref1} which has affected the results of this paper significantly. I, therefore, propose re computation of parameters after rewriting the relevant and corrected equations of Negi et al.\cite{Ref1} in Sec. 2 of the present paper by replacing the symbol $`t' \equiv `Q'$, which was assigned as  compressibility parameter in Tolman's VII solution $ (x = r^2/k^2 = r^2/K^2 = r^2/a^2t)$, discussed in Negi et al. \cite{Ref1}. Some other important properties of the models (adiabatic sound speed $({\rm d}P/{\rm d}E)^{1/2}_0$ at the centre of the star, gravitational binding energy and the pulsational stability under small radial perturbations) which were not discussed in the paper of Negi et al. \cite{Ref1} are included in Sec. 3. Results of this re computations are presented in Sec. 4. Sec. 5 summarizes the main findings obtained in this study.
\section{Matching of Parameters at the Core-Envelope Boundary}
\label{sec:2}
Rewriting expressions for energy-density corresponding to the envelope and the core regions of Negi et al.\cite{Ref1} in the following
\begin{equation}
(2n + 1 - n^2)8\pi Er^2 = n\left [(2 - n) + \frac{n(3 - n)}{(1 + n)}(r/a)^q\right ]
\end{equation}
\begin{equation}
E = E_0(1 - x); x = r^2/a^2Q
\end{equation}
the matching of energy-density $E $ at the core-envelope boundary $r = b$, that is, by setting $E_{\rm core}$ (r = b) = $E_{\rm envelope}$ (r = b), using eqs. (1) and (2) above yields the relation
\begin {equation}
(2n + 1 - n^2)8\pi E_0b^2(1 - b^2/a^2Q) = n\left [(2 - n) + \frac{n(3 - n)}{(1 + n)}(b/a)^q\right ]
\end{equation}
where $E_0$ is defined as the central energy-density and $q = 2(2n + 1 - n^2)/(1 + n)$. The continuity of $e^\nu$ and $e^\lambda$ at the surface $r = a$ yields the compactness parameter $u \equiv M/a = n/(2n + 1)$. Thus the total mass $M$ contained in the sphere is $na/(2n + 1)$, which gives
\begin{eqnarray}
\left ( \frac {na}{2n + 1}\right )& = &\int_{0}^{b} 4\pi E_{\rm core}r^2 dr + \int_{b}^{a} 4\pi E_{\rm envelope}r^2 dr \nonumber \\
& = & I_1 + I_2
\end{eqnarray}
Using eqs. (2) and (1) above, $I_1$ and $I_2$ may be evaluated as
\begin{equation}
I_1 = (4\pi E_0b^2/3)a[1 - (3b^2/5a^2Q)](b/a)
\end{equation}
(note that the numeral $3$ is missing in numerator of the second term of eq.(21) corresponding to $I_1$ of Negi et al.\cite{Ref1}.)
\begin{equation}
I_2 = [na/2(2n + 1 -n^2)] \left [ (2 - n)[1 - (b/a)] + \frac{n}{2n + 1}[1 - (b/a)^{q + 1}] \right ]
\end{equation}
Combining eqs.(3) - (6), we get
\begin{equation}
8\pi E_0b^2 = 6 \left[ \frac{n}{(2n + 1)} - \frac{I_2}{a}\right]\left[ \left(1 - \frac{3b^2}{5a^2Q}\right)\frac{b}{a}\right ]^{-1}
\end{equation}
where $8\pi E_0b^2$ and $I_2$ are given by eqs.(3) and (6). For a given $n$ and $Q$ values, one can obtain a $(b/a)$ value for which eq.(7) is satisfied. This ensures the matching of $E$ and $\lambda$ (and also $\rm d\lambda/\rm {d r}$) at $r = b$.
I have re investigated that this matching can be ensured for the values of $n$ in the range $0 < n \leq (3/4)$ and the matching does not exists for the values of $n > (3/4)$. I have carried out in the present study, this matching for the values of $n = (1/2)$ and $n = (3/4)$ [that is for $u$ values  0.25 and 0.30 respectively]. The matching for other allowed values of $n$ in the range prescribed above can also be done likewise.

As soon as the value of $(b/a)$ is obtained by using eq.(7) above, one can also calculate the value of $(8\pi E_0b^2)$ by using eq. (3). Now substituting $x = r^2/a^2Q$ in eq.(16) of Negi et al. \cite{Ref1} and rewriting the expression for $w$ as
\begin{eqnarray}
w & = & \rm{ln} \{ (r^2/a^2Q) - (5/6) \nonumber \\
& & + \left [ ( r^2/a^2Q)^2 - (5r^2/3a^2Q)+(5/8\pi E_0a^2Q) \right ]^{1/2} \}
\end{eqnarray}

Substituting the value of $(8\pi E_0b^2)$ in eq.(8) above, the value of $w_b$ (the value of $w$ at $r = b$) can be calculated in the following form (note that the parameter $Q$ is missing in denominator of the fourth term in expression of $w_b$ of Negi et al.\cite{Ref1})
\begin{eqnarray}
w_b & = & \rm{ln} \{ (b^2/a^2Q) - (5/6) \nonumber \\
& & + \left [ ( b^2/a^2Q)^2 - (5b^2/3a^2Q)+(5/8\pi E_0b^2)(b^2/a^2Q) \right ]^{1/2} \}
\end{eqnarray}
Now rewriting eqs.(13) and (6) for $e^\nu$ and eqs.(15) and (11) for pressure $P$ corresponding to the core and the envelope regions of Negi et al. \cite{Ref1} as 
\begin{equation}
e^{\nu/2} = C_1{\rm cos}(w/2) + C_2{\rm sin}(w/2)
\end{equation}
\begin{equation}
e^\nu = \frac{1}{(2n + 1)}(r/a)^{2n}
\end{equation}
\begin{equation}
8\pi P = M\frac{C_2{\rm cos}(w/2) - C_1{\rm sin}(w/2)}{ C_1{\rm cos}(w/2) + C_2{\rm sin}(w/2)} - N
\end{equation}

\begin{equation}
(2n + 1 - n^2)8\pi Pr^2 = n^2[1 - (r/a)^q]
\end{equation}

By using eqs.(10) and (11) and (12) and (13) in pairs, I match $\nu$ and $P$ (and also $\rm d\nu/\rm {dr}$) at $r = b$. By setting $\nu_{\rm core} (r = b) = \nu_{\rm envelope} (r = b)$ and $P_{\rm core} (r = b) = P_{\rm envelope} (r = b)$, I obtain $C_1$ and $C_2$ from the relevant equations given in Negi et al. \cite{Ref1} as
\begin{equation}
C_1 =  A_1{\rm cos}(w_b/2) -  B_1{\rm sin}(w_b/2)
\end{equation}
\begin{equation}
C_2 =  A_1{\rm sin}(w_b/2) +  B_1{\rm cos}(w_b/2)
\end{equation}
where
\begin{equation}
A_1 =  (b/a)^n/(2n + 1)^{1/2}
\end{equation}
\begin{equation}
B_1 =  (A_1/M_bb^2) \left ( \frac{n^2}{(2n + 1 - n^2)} \left [ 1 -  (b/a)^q \right ] + N_bb^2 \right )
\end{equation}
and
\begin{eqnarray}
M_bb^2 & = &  \left ( \frac{32\pi E_0b^2(b/a)^2}{5Q}\right )^{1/2} \nonumber \\
                     && \times  \left [ 1 - \frac{(8\pi E_0b^2)}{15(b^2/a^2Q)}  [5(b^2/a^2Q) - 3(b^2/a^2Q)^2 ] \right ]^{1/2} 
\end{eqnarray}
\begin{equation}
N_bb^2 =  \left ( \frac{8\pi E_0b^2}{15} \right )[5 - 3(b^2/a^2Q)]
\end{equation}
Having calculated $w_b, (8\pi E_0b^2), C_1$ and $C_2, P, E, \lambda$ and $\nu$ are known throughout the configuration. Furthermore, by using eqs. (2) and (1) above the ratios $(E_0/E_a)$ and $(E_0/E_b)$ may also be calculated. Finally, by assigning the surface density $E_a$ to be equal to that of the average nuclear density ($2 \times 10^{14}\rm {g cm}^{-3}$, \cite{Ref10}) the mass and size (radius) of neutron star models based on the present study can be calculated.
\section{Gravitational Binding and Pulsational Stability of Core-Envelope Models}
\label{sec:3}
The coefficient of gravitational binding $\alpha_{gb}$ and the ratio of gravitational packing $\alpha_p$ can be obtained by using the equations \cite{Ref11}
\begin{equation}
\alpha_{gb} = \alpha = (M_r - M)/M_r = [(M_r/a) - (M/a)]/(M_r/a)
\end{equation}

\begin{equation}
\alpha_p = (M_p - M)/M = [(M_p/a) - (M/a)]/(M/a)
\end{equation}
where $M/a, M_r/a$ and $M_p/a$ are given by the relations
\begin{equation}
M/a = 4\pi \int_{0}^{1}Ea^2y^2dy
\end{equation}
\begin{equation}
M_r/a = 4\pi \int_{0}^{1}\rho a^2y^2 e^{\lambda/2}dy
\end{equation}
\begin{equation}
M_p/a = 4\pi \int_{0}^{1}Ea^2y^2 e^{\lambda/2}dy
\end{equation}
where $\rho = (P + E) e^{(\nu - \nu_a)/2}$ is called the rest-mass density (Durgapal \& Pande \cite{Ref12})and $y \equiv r/a$ is the radial coordinate measured in units of configuration size.

The pulsational stability of the structures under small radial perturbations can be judged by using variational method \cite{Ref13}. For a stable configuration the pulsational frequency is given by
\begin{equation}
f = (1/2\pi)(A/B)^{1/2}
\end{equation}
where the functions $A$ and $B$ are respectively the potential energy and the kinetic energy with velocities replaced by displacements and are given by{\footnote 
{For simplification these expressions are obtained by using the `trial function' $\xi = re^{\nu/2}$, because this trial function is sufficient to judge the pulsational stability as obtained by using the trial function of the form of a power series ( \cite{Ref14}; and references therein) $\xi = b_1r(1 + a_1r^2 + a_2r^4 + a_3r^6)e^{\nu/2}$, where $a_1, a_2,$ and $a_3$ are arbitrary constants. Furthermore, the study of Knutsen \cite{Ref15} also shows that the use of the trial function of the form of the power series mentioned above (with suitable values of the arbitrary constants $a_1, a_2,$ and $a_3$ such that the appropriate boundary conditions may be satisfied) provide the results similar to those obtained by using the trial function $\xi = re^{\nu/2}$.}}
\begin{equation}
8\pi B/a^3 = \int_{0}^{1}(8\pi Pa^2 + 8\pi Ea^2)y^4e^{(3\lambda + \nu)/2}dy
\end{equation}
and
\begin{eqnarray}
8\pi A/a& = &\int_{0}^{1}y^2e^{3(\lambda + \nu)/2}\{ e^{-\lambda}[9(8\pi Pa^2 + 8\pi Ea^2)(dP/dE) \nonumber \\
& & + 4(8\pi a^2dP/dy)y - \frac{(8\pi a^2dP/dy)^2y^2}{(8\pi Pa^2 + 8\pi Ea^2)}] \nonumber \\
& & + 8\pi Pa^2(8\pi Pa^2 + 8\pi Ea^2)y^2 \}dy
\end{eqnarray}
Eqs. (26) and (27) may be computed by employing a fourth order Runge-Kutta method from the centre ($y = 0$) to the boundary ($y = b/a$) by using Tolman's VII solution and from the boundary ($y = b/a$) to the surface ($y = 1$) by using Tolman's V solution which yield the values of function $(8\pi B/a^3)$ and $(8\pi a/A)$. On dividing values obtained by using eq.(27) by eq.(26) one gets the value of $a\omega$, where $\omega$ being the angular frequency of pulsation which follows from eq.(25). On computation, the positive values of pulsation frequencies would show that the average (constant) value of adiabatic index, $\gamma_{\rm ave}$, is larger than the minimum (critical) value of (constant) adiabatic index, $\gamma_{\rm crit}$, required for the stability of the structures (that is, $\gamma_{\rm ave} \geq \gamma_{\rm crit}$). Thus, we can safely conclude that the structures are stable under small radial perturbations. This is to be pointed out here that the use of the trial function $\xi = re^{\nu/2}$ in the above eqs. (26) and (27) safely assures the pulsational stability of the models considered in this study, because the present models correspond to the value of $u < 1/3$. For $u \geq 1/3$, the optimal trial function $\xi = re^{\nu/4}$ may be used for ascertaining the pulsational stability (see, e.g.\cite{Ref16}, \cite{Ref9}) which is not required in the present study.
The various variables appear in eqs. (20) - (27) are given in Negi et al. \cite{Ref1}, however some additional variables which are not given in Negi et al. \cite{Ref1} and defined in the present paper are given below

\subsection{The Core: $0 \leq y \leq (b/a)$}
\label{sec:1}
\begin{equation}
-8\pi a^2 \frac{dE}{dy} = 8\pi E_0a^2(2y/Q)
\end{equation}

\begin{equation}
-8\pi a^2 \frac{dP}{dy} =(1/2)( 8\pi Pa^2 + 8\pi Ea^2)(8\pi Pa^2y^2 + 1 - e^{-\lambda})e^\lambda y^{-1}
\end{equation}
\begin{equation}
8\pi \rho a^2 = (8\pi Pa^2 + 8\pi Ea^2)e^{\nu/2}(1 - 2u)^{-1/2}
\end{equation}

\subsection{The Envelope: $(b/a) \leq y \leq 1$}
\label{sec:2}
\begin{eqnarray}
-8\pi a^2 \frac{dE}{dy}y^3 & = & [2n/(2n + 1 - n^2)] \nonumber \\
& & \times \{ (2 - n) -[n^2(3 - n)(1 - n)y^q/(n + 1)^2]y^q\}
\end{eqnarray}

\begin{equation}
-8\pi a^2 \frac{dP}{dy} =(1/2)( 8\pi Pa^2 + 8\pi Ea^2)(8\pi Pa^2y^2 + 1 - e^{-\lambda})e^\lambda y^{-1}
\end{equation}
\begin{equation}
8\pi \rho a^2 = (8\pi Pa^2 + 8\pi Ea^2)e^{\nu/2}(1 - 2u)^{-1/2}
\end{equation}
\section{Results}
\label{sec:4}
The variation of $(b/a)$ with $Q$ for $n = 1/2 (u = 0.25)$ and 3/4 ($u = 0.30)$ is shown in Table 1 and Table 2. As $Q$ increases, so does $(b/a)$. At a certain $Q$ value, $(b/a)$ becomes equal to 1. For this optimum $Q$ value (e.g. $n = (3/4), (b/a) \simeq 1$ at $Q = 1.3$), the entire configuration corresponds to TDR - solution. As $Q \rightarrow 0$, $(b/a) \rightarrow 0$ and the entire configuration pertains to Tolman's V solution. The density ratios have been computed for $ n = (1/2)$ and (3/4) and the results are shown in Table 1 and Table 2 respectively. As $Q$ increases both $(E_0/E_b)$ and $(E_0/E_a)$ follow a decreasing trend. As $b \rightarrow a$, the ratios tend to become equal. For $Q \rightarrow 0$, $E_0 \rightarrow \infty$ (Tolman's V solution).

The surface redshift depends only on the $n$ value. The boundary redshift may be calculated straight away from eq.(6) of Negi et al. \cite{Ref1}. The central redshift, $z_0$, however is calculated by using eq. (13) of Negi et al.\cite{Ref1}. The variation of $z_0$ with $Q$ for $n = (1/2)$ and (3/4) is also given in Table 1 and Table 2 respectively. It is seen that $z_0$ increases quite rapidly with decreasing $Q$ and as $Q \rightarrow 0$, $z_0 \rightarrow \infty$.

For these calculations $E_a$ has been taken to be $2 \times 10^{14}\rm {g cm}^{-3}$ ( like, Brecher \& Caporaso \cite{Ref10}). Because for the models considered in the present study, the speed of sound, $v_s$, remains finite and significantly less than the speed of light in vacuum, $c = 1$, at the surface where pressure vanishes.Therefore, it seems physically plausible to assume that the matter represents a self-bound state at the surface density of average nuclei ($E_a = 2\times 10^{14}\rm g{cm}^{-3}$ ). This feature is similar to the models corresponding to the EOSs of quark stars where pressure vanishes at the finite surface density (see, e.g.\cite{Ref17}, \cite{Ref18}; and references therein). The total size of the configuration depends only on $n$ value and turns out to be 13.369 km and 15.157 km for $n = (1/2)$ and $n = (3/4)$ respectively. The core size depends also on the value of $Q$ together with $n$. For $Q = 0.1$ the core radii have the values 3.850 km and 4.350 km respectively for the two cases $n = (1/2)$ and $n = (3/4)$. The masses of the models depend only on $n$ and have the values 2.267$M_\odot$ and 3.085$M_\odot$ for $n = (1/2)$ and $n = (3/4)$ respectively.

The variation of central pressure, $P_0$, with $Q$ can also be calculated by using eq. (15) of Negi et al. \cite{Ref1}. Table 1 and Table 2 show the variation of central pressure, $P_0$, with $Q$ for $n$ values (1/2) and (3/4) respectively. In both the cases as $Q$ decreases, i.e. as the core size decreases, $P_0$ increases quite rapidly and as $Q \rightarrow 0$, $P_0 \rightarrow \infty$ corresponding to a singularity at the centre in the Tolman's V solution.

The variation of the ratio of central pressure to central energy-density, $(P_0/E_0)$, with $Q$ is shown is Fig.1 for $n = (1/2)$ and $n = (3/4)$ respectively. For Tolman's V solution the value of $(P_0/E_0)$ becomes (1/3) when $n = (1/2)$; but in the present model $(P_0/E_0) = 0.348$ for $n = (1/2)$ and $Q = 0.001$. For $n = (3/4)$ the present model yields $(P_0/E_0) = 0.620$ and $Q = 0.001$. This feature is common among realistic models of neutron stars available in the literature.

The value of ${\rm d} P/{\rm d} E$ which represents the square of adiabatic sound speed ($v_s \equiv ({\rm d} P/{\rm d} E)^{1/2}$) has also been calculated. Fig. 2 shows variation of ${\rm d} P/{\rm d} E$ at $r = 0$, i.e. $({\rm d} P/{\rm d} E)_0$ with varying $Q$ for $u$ values (1/2) and (3/4). It is seen that $({\rm d} P/{\rm d} E)_0 < 1$ (that is , the causality condition is fulfilled) and it decreases slightly as $Q$ changes from 0.001 to 1.3.

The binding energy coefficients, $\alpha_{gb}$ and $\alpha_p$, of the models considered in the present study are shown is Fig.3 and Fig. 4 for $n$ values (1/2) and (3/4) respectively. The values of $\alpha_{gb}$ and $\alpha_p$ indicate that the structures are gravitationally bound for all possible values of $Q$ and $u$. As $Q \rightarrow 1.3$ the values of $\alpha_{gb}$ and $\alpha_p$ become closer to each other. However, this value of $Q$ corresponds to a slower variation of density inside the structure (which corresponds to a structure with a negligible envelope, i.e. the entire configuration is represented by Tolman's VII solution) so much so that the rest mass density and the energy density become almost equal. Furthermore, it may be noted that $\alpha_{gb}$ is continuously increasing with $Q$ for both the values of $u$ = 0.25 and $u$ = 0.30 which means that the structures are also pulsationally stable together with the property that they are gravitationally bound which is the outcome of binding energy criterion of fluid stars which states that the configurations remain pulsationally stable upto the first maxima in the binding energy curve \cite{Ref11}, \cite{Ref19}.

Fig. 5 gives a plot between $a\omega$ and $Q$ for $u = 0.25$ and $u = 0.30$ respectively. The positive values of $a\omega$ indicate that the structure is pulsationally stable for both the values of $u$ considered in the present study.
\section{Summary}
\label{sec:5}
A massive configuration corresponding to a core described by TDR-Solution and the envelope is given by Tolman's V solution has been re investigated and the new calculations for various important physical properties have been provided. The study describes a model for which all the four variables $P, E , \nu$ and $\lambda$ along with ($\rm d\nu/\rm dr$) and ($\rm d\lambda/\rm dr$) are continuous at the core-envelope boundary $r = b$.

The model is causal, gravitationally bound and pulsationally stable and corresponds to an upper bound on neutron star mass, $M \simeq 3.085 M\odot$, which represents the mean value of the classical result of maximum mass, $M \simeq 3.2 M\odot$ obtained by Rhodes \& Ruffini \cite{Ref3} and the result of the secure upper bound on neutron star mass $M \simeq 2.9 M\odot$
obtained by Kalogera \& Byam \cite{Ref4} on the basis of modern EOSs for neutron star matter. The maximum mass obtained in this study, however, is much closer to the maximum mass obtained recently by Sotani \cite{Ref5}. Furthermore, the observational constraint imposed by the recently measured largest pulsar masses around 2$M_\odot$ \cite{Ref6}, \cite{Ref7}, \cite{Ref8} is comfortably satisfied by the models considered in the present study. The maximum allowed value of compactness parameter $u \leq 0.30$ obtained in this study is also consistent with an absolute maximum value of $u_{\rm max} = 0.333^{+0.001}_{-0.005}$ resulting from the combination of results obtained by Bauswein et al.\cite{Ref20} and Margalit and Metzger\cite{Ref21} from the observation of binary neutron stars merger GW170817  (see, e.g.\cite{Ref9}).

\section{Acknowledgments}

 The author is grateful to anonymous referee for his valuable comments, suggestions and comprehensive reviewing of the present paper.


\section {Conflict of Interest Statement:} The Author declares that there is no conflict of interest.

\begin{figure*}
\includegraphics[width=0.75\textwidth]{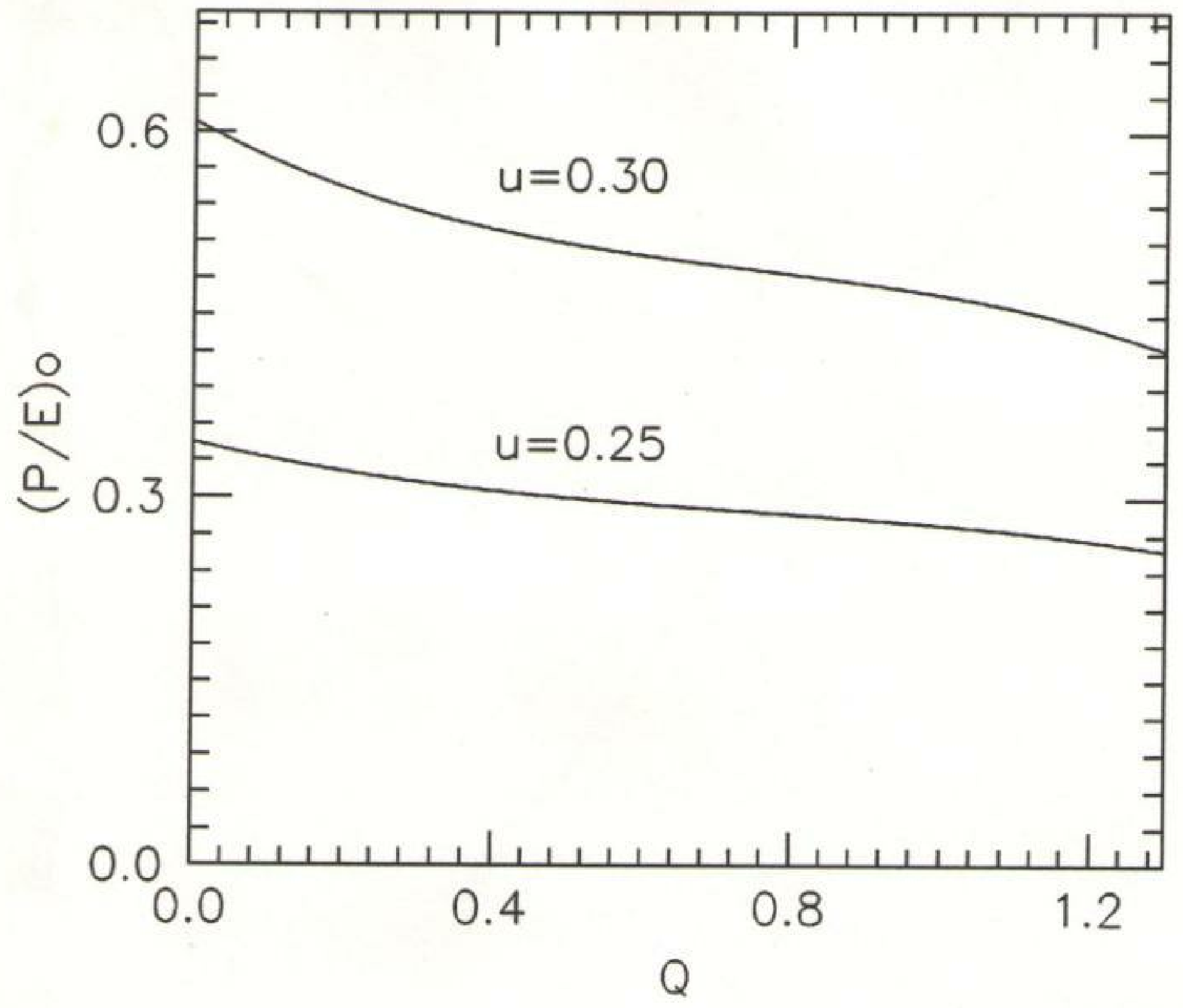}
\caption{Variation of $(P/E)_0$ with $Q$ for $u$ values 0.25 and 0.30.}
\label{fig:1} 
\end{figure*}

\begin{figure*}
\includegraphics[width=0.75\textwidth]{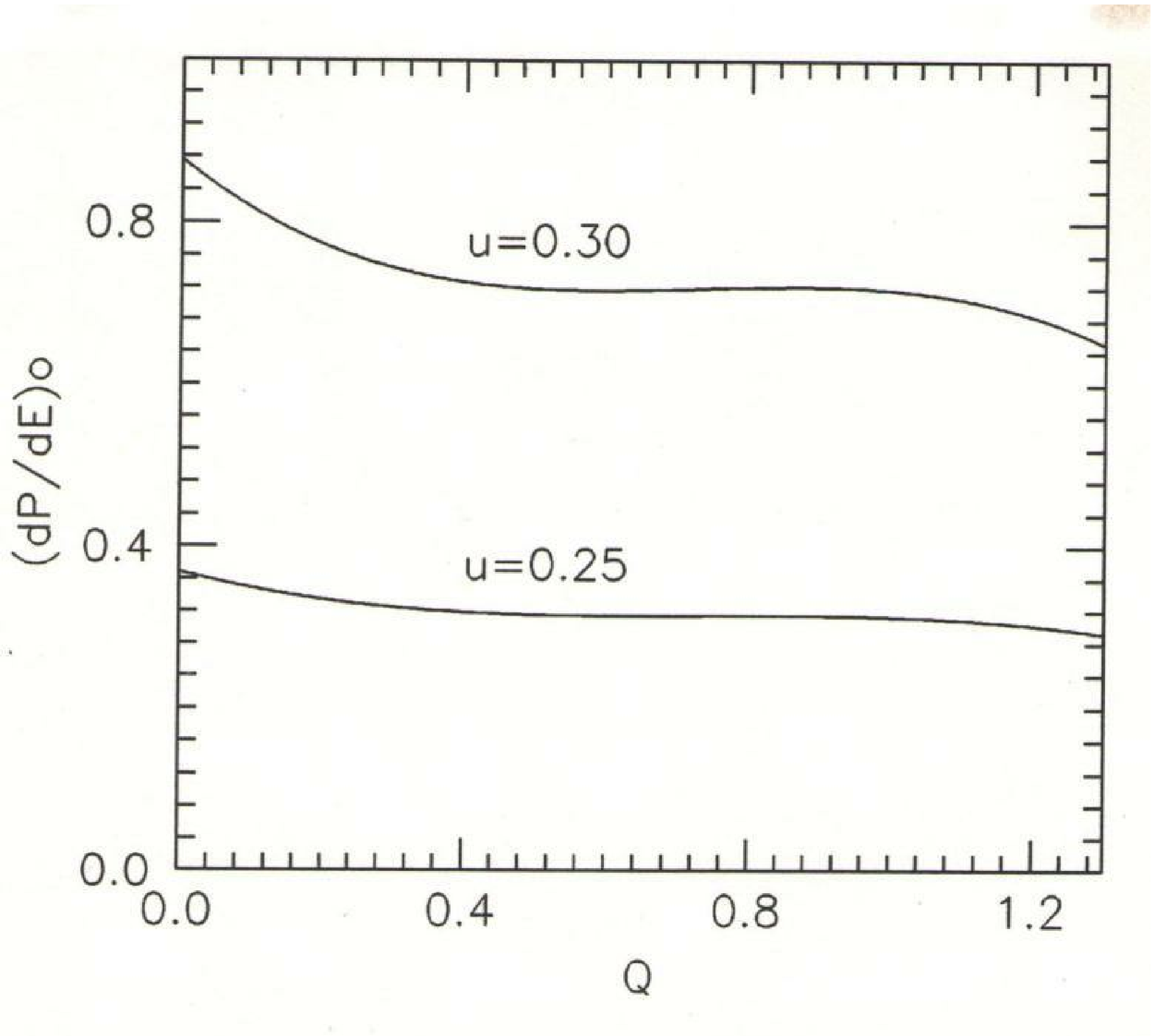}
\caption{Variation of $(\rm d P/\rm d E)_0$ with $Q$ for $u$ values 0.25 and 0.30.}
\label{fig:2} 
\end{figure*}

\begin{figure*}
\includegraphics[width=0.75\textwidth]{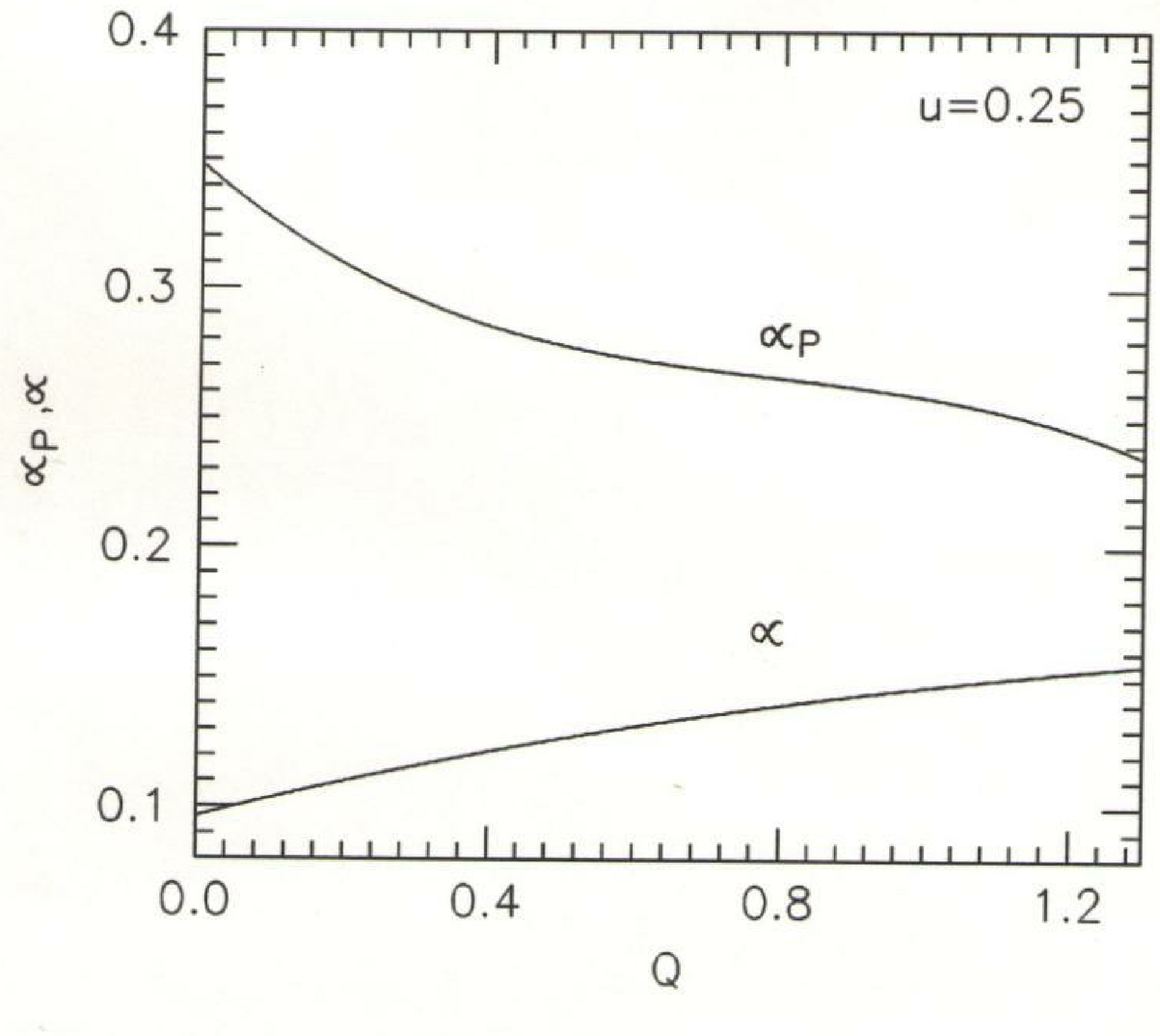}
\caption{Variation of $\alpha = \alpha_{gb}, \alpha_p$ with $Q$ for the values of $u = 0.25$.}
\label{fig:3} 
\end{figure*}

\begin{figure*}
\includegraphics[width=0.75\textwidth]{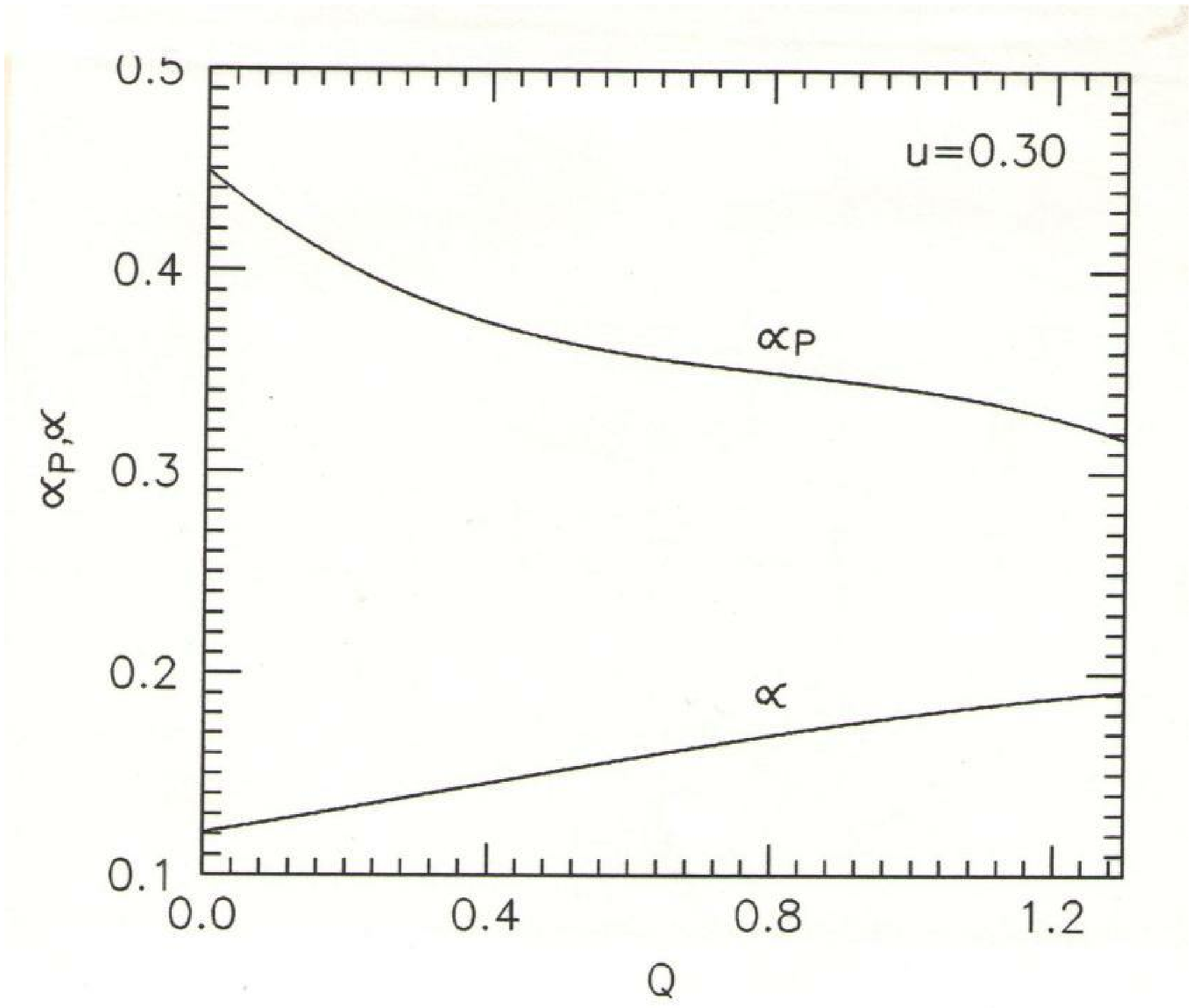}
\caption{Variation of $\alpha = \alpha_{gb}, \alpha_p$ with $Q$ for the values of $u = 0.30$.}
\label{fig:4} 
\end{figure*}

\begin{figure*}
\includegraphics[width=0.75\textwidth]{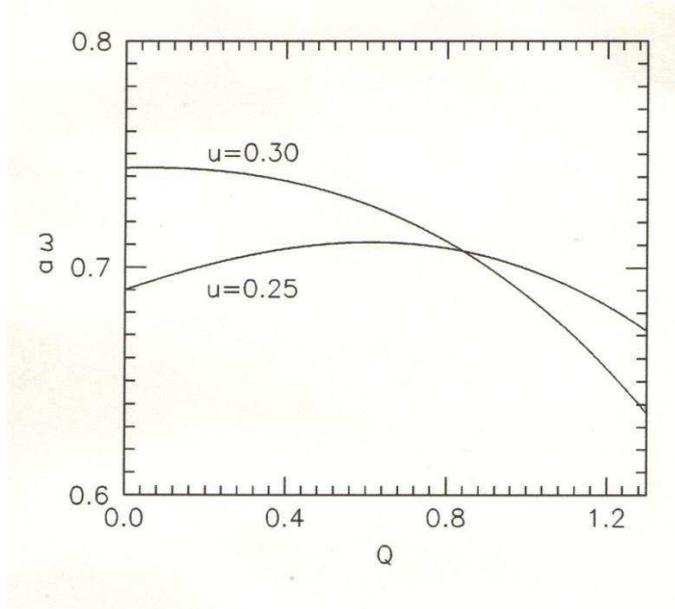}
\caption{Variation of $a\omega$ with $Q$ for $u$ values 0.25 and 0.30.}
\label{fig:5} 
\end{figure*}
%
\begin{table}
\caption{The Core-Envelope Boundary, $(b/a)$, Total Radius, $a$, Core Radius, $b$, Central Energy- Density, $E_0$, Boundary and Surface values of Energy-Density ($E_b$ and $E_a$),  ratios of Central Energy-Density to the Boundary and the Surface Energy-Density  ($E_0/E_b$ and  $E_0/E_a$), Central Red-shift, $z_0$, and Central Pressure, $P_0$ for an assigned value of $n = 1/2 (u = 0.25)$ and various allowed values of $Q$. The value of Surface Density, $E_a$, is assumed to be the average nuclear density, like Brecher \& Caporaso (\cite {Ref10}). For these values of $u$ and $E_a$ the total mass of the configuration corresponds to a value of 2.267$M_\odot$ (where$M_\odot \simeq 1.474$km).}
\label{tab:1} 
\begin{tabular}{llllllllllll}
\hline\noalign{\smallskip}
& & & & & & $ u = 0.25$ & & & & & \\
\hline\noalign{\smallskip}
$Q$ & $(b/a)$ &$ a$(km) &$ b$(km) & $8\pi E_0b^2$ & $8\pi E_bb^2$ & $E_0/E_b$ & $8\pi E_0a^2$ & $8\pi E_aa^2$ & $E_0/E_a$ & $z_0$ & $8\pi P_0a^2$ \\
\noalign{\smallskip}\hline\noalign{\smallskip}
0.001 & 0.029 & 13.369 & 0.388 & 2.696 & 0.429 &6.284 & 3205.708 & 0.666 & 4813.375 & 11.048 & 1114.285 \\
0.100 & 0.288 & 13.369 & 3.850 & 2.589 & 0.442 & 5.857 & 31.214 & 0.666 & 46.868 & 2.759 & 10.204 \\
0.300 & 0.495 & 13.369 & 6.618 & 2.590 & 0.475 & 5.453 & 10.570 & 0.666 & 15.871 & 1.849 & 3.309 \\
0.600 & 0.693 & 13.369 & 9.265 & 2.654 & 0.530 & 5.008 & 5.526 & 0.666 & 8.297 & 1.410 & 1.650 \\
0.900 & 0.839 & 13.369 & 11.216 & 2.693 & 0.587& 4.588 & 3.826 & 0.666 & 5.745 & 1.183 & 1.074 \\
1.300 & 0.993 & 13.369 & 13.275 & 2.744 & 0.663 & 4.139 & 2.783 & 0.666 & 4.179 & 1.000 & 0.717 \\
\noalign{\smallskip}\hline\noalign{\smallskip}
\end{tabular}
\end{table}

\begin{table}
\caption{The Core-Envelope Boundary, $(b/a)$, Total Radius, $a$, Core Radius, $b$, Central Energy-Density, $E_0$,  Boundary and Surface values of Energy-Density ($E_b$ and $E_a$),  ratios of Central Energy-Density to the Boundary and the Surface Energy-Density  ($E_0/E_b$ and  $E_0/E_a$), Central Red-shift, $z_0$, and Central Pressure, $P_0$ for an assigned value of $n = 3/4 (u = 0.30)$ and various allowed values of $Q$. The value of Surface Density, $E_a$, is assumed to be the average nuclear density, like Brecher \& Caporaso (\cite {Ref10}). For these values of $u$ and $E_a$ the total mass of the configuration corresponds to a value of 3.085$M_\odot$ (where$M_\odot \simeq 1.474$km).}
\label{tab:2}
\begin{tabular}{llllllllllll}
\hline\noalign{\smallskip}
& & & & & & $ u = 0.30$ & & & & & \\
\hline\noalign{\smallskip}
$Q$ & $(b/a)$ &$ a$(km) &$ b$(km) & $8\pi E_0b^2$ & $8\pi E_bb^2$ & $E_0/E_b$ & $8\pi E_0a^2$ & $8\pi E_aa^2$ & $E_0/E_a$ & $z_0$ & $8\pi P_0a^2$ \\
\noalign{\smallskip}\hline\noalign{\smallskip}
0.001 & 0.029 & 15.157 & 0.440 & 3.044 & 0.484 & 6.289 & 3619.501 & 0.857 & 4223.455 & 39.000 & 2244.561 \\
0.100 & 0.287 & 15.157 & 4.350 & 2.878 & 0.507 & 5.676 & 34.940 & 0.857 & 40.770 & 5.757 & 19.483 \\
0.300 & 0.493 & 15.157 & 7.472 & 2.960 & 0.562 & 5.267 & 12.179 & 0.857 & 14.211 & 3.545 & 6.610 \\
0.600 & 0.687 & 15.157 & 10.413 & 3.029 & 0.646 & 4.689 & 6.418 & 0.857 & 7.489 & 2.521 & 3.231 \\
0.900 & 0.830 & 15.157 & 12.580 & 3.116 & 0.731 & 4.263 & 4.523& 0.857 & 5.278 & 2.058 & 2.136 \\
1.300 & 0.979 & 15.157 & 14.839 & 3.197 & 0.840 & 3.806 & 3.336 & 0.857 & 3.893 & 1.674 & 1.411 \\
\noalign{\smallskip}\hline\noalign{\smallskip}
\end{tabular}
\end{table}




%
%


\end{document}